# Forensic Statistics and Justice: the Leiden Consensus

Ian Freckelton[1], Richard D. Gill[2], and Johannes F. Nijboer[3]

This version: 15 March, 2012

**Background**

Between 26 and 29 April 2011 a four-day interdisciplinary workshop "Science meets Law" was held at the Lorentz Center (University of Leiden, NL). It was organised by the Lorentz Center in collaboration with the Netherlands Institute for Advanced Study in the Humanities and Social Sciences ("NIAS", Wassenaar, NL). The programme was prepared by Richard Gill and Johannes F. Nijboer and connected to Richard Gill's 2010-2011 stay at NIAS as Distinguished Lorentz Fellow.

Hans Nijboer passed away 13 April, 2013; age 61. We miss a good friend and a great scientist.

**A Time of Major Change and Challenge for Forensic Science and for the Development of the Law**

Forensic science is undergoing a period of fundamental change.

In common law countries there have recently been dramatic developments: the imminent privatisation of the Forensic Science Service (FSS) in England; authorisation of disciplinary hearings against registered forensic experts in the United Kingdom (*Meadow v General Medical Council, 2006*); abolition of the civil immunity of expert witnesses in the United Kingdom (*Jones v Kaney*, 2011); likely changes to the admissibility of opinions in the United States (Amendments to the Federal Rules of Evidence, 2011) and the United Kingdom (Law Commission of England and Wales, 2011), after major changes to statutory regimes for evidentiary admissibility in Australia and New Zealand (see Freckelton, 2009); and new scrutiny upon the way in which expert opinions should be expressed to the courts.

In Continental criminal justice systems there is also heightened attention for the weaknesses and risks of poor quality forensic science. Within the European Union the mutual recognition of "products" of investigation, prosecution and adjudication has been given high priority (exchange of data from databases on the basis of the Prüm treaty and the subsequent regulation by the European Commission, the so-called Swedish framework decision on the accreditation and standardisation in the fields of DNA-related and biometric forensic expertise). In Belgium and the Netherlands initiatives have been launched for quality improvement and control: see the Belgian report by the Hoge Raad voor de Justitie/Conseil Supérieur de la Justice of 30 March 2011, concerning the statute and quality of "court experts" (gerechtsdeskundigen) and the Dutch legislation on the position of experts in criminal proceedings (Wet Deskundige in strafzaken, in force 1 January 2010) and the related creation of a national register (Nederlands Register Gerechtelijk Deskundigen, NRGD). In many countries the discovery of judicial errors (miscarriages of justice), sometimes closely related to the involvement of forensic expertise in the case, has given a boost to more intensive attention to this aspect of the criminal justice system (De Roos & Nijboer, 2011).

---

[1] Senior Counsel, Victorian Bar, Australia; Professor of Law, Forensic Medicine and Forensic Psychology, Monash University; I.Freckelton@vicbar.com.au

[2] Mathematical Institute, University Leiden; gill@math.leidenuniv.nl

[3] Law Faculty, University Leiden; j.f.nijboer@law.leidenuniv.nl

These phenomena will have important flow-on effects for the investigation and research carried out by and the evidence given by forensic statisticians and other forensic scientists. They provide important opportunities for the disciplines to gain enhanced standing and credibility in the courts and in the criminal justice system. In theory this pertains for all jurisdictions in which non-legal (forensic) expertise plays a significant role, including the activities of special criminal tribunals (like the International Criminal Tribunal for the former Yugoslavia ) and of the International Criminal Court.

**Courts' Discomfort with Numerate Evidence**

Many (criminal) court decisions have expressed discomfort about the potential for evidence expressed in numerate terms to be misconstrued by both judges and juries – this was the basis for the preclusion upon the *"prosecutors' fallacy"* (see Thompson and Schumann, 1987). In 1997 in *Doheny and Adams*, for instance, the English Court of Appeal strongly endorsed the observation in *Adams* (1992: 482) that:

"To introduce Bayes Theorem, or any similar method, into a criminal trial plunges the jury into inappropriate and unnecessary realms of theory, and complexity deflecting them from their proper task."

While the quest for certainty remains (see Vosk, 2010), at least to the point of evaluating whether evidence proves guilt beyond reasonable doubt, the potential of "trial by numbers" remains a source of fear for courts in many jurisdictions.

Four years after *Doheny* in Australia the then President of the New South Wales Court of Criminal Appeal (Mason P in *R v GK* (2001) at [26]–[27]) expressed similar sentiments in emphatic terms: The process of assessing the weight of different items of evidence and reasoning to a conclusion on the civil or criminal standard cannot be reduced to mathematical formulae. In an article on "Probability and Proof in Legal Fact Reasoning" (1995) 69 ALJ 731 at p736, the Hon Mr Justice D H Hodgson said that *"decision-making generally involves a global assessment of a whole complex array of matters which cannot be given individual numerical expression"*. He warned that concentration on mathematical probabilities can prejudice the common sense process which depends upon experience of the world and belief as to how people generally behave (see also *State Government Insurance Commission v Laube* (1984) 37 SASR 31 at 32-3, *Mitchell* (1997) 98 A Crim R 32 at 37-8, *Burger King Corporation v Hungry Jack's Pty Ltd* [2001] NSWCA 187 at [591]).

It is therefore inappropriate to determine guilt by the application of mathematical formulae suggesting how to aggregate the impact of different items of evidence. That is why Bayes Theorem has been rejected in this area.

Judicial officers generally have little background in numeracy and have not adapted readily to evidence in a quantitative form – both because of a concern that they or, especially, a lay jury, may misconstrue it, and because of the risk that it may overwhelm and distort the decision-making process in both criminal and civil hearings. As Lord Kerr put it in *Sienkiewicz v Grief (UK) Ltd* (2011: at [2011]):
There is a real danger that so-called 'epidemiological evidence' will carry a false air of authority. It is necessary to guard against treating a theory based on assumptions as a workable benchmark against which an estimate of the increase in risk could be measured. (see generally Goldberg, 2011)

In the now notorious decision of *R v T* (2010; see Redmayne, Roberts, Aitken, and Jackson, 2011), part of the difficulty from the perspective of the Court of Appeal of England and Wales was that for good reasons, which were clearly explained on appeal, the scientist involved elected not to present his opinions in a relative risk ratio or in a numerate form (endeavouring to refrain from any form of prosecutor's fallacy), rather undertaking careful calculations and then expressing his views in a qualitative, generalised way (cp *R v South*, 2011). The anxiety expressed by the Curt of Appeal in relation to anything resonant of quantitative analysis is representative of the law's traditional inclination toward decision-making flexibility and qualitative analysis of *"all of the evidence"* at trial and the adoption of an approach of unfettered *"intuitive synthesis"* of relevant factors at sentencing.

A difficulty arises too for courts in combining quantitative with qualitative evidence in a rigorous way. An aspect of this is the disentanglement of "hard" evidence from "soft", just as courts have bewailed the problems that arise when insufficient distinction is drawn between evidence in the form of "fact" and evidence in the form of "opinion", as well as in relation to inadequately articulated assumptions made on the way to the formation of an opinion (see eg *Makita (Australia) Pty Ltd v Sprowles,* 2001: at [85]; see also *Dasreef Pty Ltd v Hawchar,* 2011).

A challenge, therefore, for forensic statisticians arises in relation to how they can shift a culture of mistrust and discomfort on the part of judicial officers towards acceptance of the contribution that can be made from a quantitative perspective to the decision-making process without overwhelming or distorting it.

At present forensic statisticians world-wide are trying to present their conclusions in qualitative terms intended to be suitable for *unfettered intuitive synthesis of all of the evidence*. There is a strong consensus within the professional community that the forensic statistician should report a so-called *likelihood ratio*: that is to say, the ratio of the probabilities of the statistical evidence in question, under hypotheses appropriate to the two parties in a criminal case: defence and prosecution. This does not oblige the court to combine this part of the evidence with other parts in a formal application of Bayes' rule. A likelihood ratio of, say, 2000, can be characterised as *"strong support for the proposition"* according to the *"Standards for the formulation of evaluative forensic science expert opinion"* proposed by the *Association of Forensic Science Providers* (2009).

It is very clear that the legal community is still far from understanding adequately the enormous difference between proposals that courts should adopt Bayes' rule and formal probability calculus to come to their conclusions, and that courts should be prepared to accept forensic statistical evidence presented in the form of likelihood ratios. This is certainly not helped by forensic statisticians' own use of language, which often clouds the distinction.

On the other hand, it is reasonable that courts should be wary of *a false air of authority* carried by results of quantitative analysis. Likelihood ratios have precise meanings and typically depend on a large number of assumptions, some relatively *hard* but some extremely *soft*. Here the forensic statistical community has a great deal of work to do in developing means to communicate the reliability and the domain of validity of a particular analysis.

A likelihood ratio is based on a mathematical/probabilistic model and formulation of the problem to be considered. This model (or rather pair of models) is a necessarily *idealised*, hence necessarily *highly simplified*, description of the situation. The important question, which the statistical expert is responsible for answering in discussion with the court, is whether or not it is still adequate for the *purpose at hand*. The model should be as simple as possible, without prejudicing its adequacy, in order to be as clearly transparent as possible. Whether or not these conflicting aims (*transparency* and *realism*) can be satisfied at the same time (i.e., without sacrificing *adequacy*) can never be a prior guaranteed. There are a whole host of issues about sensitivity of model assumptions (e.g. do we account for various possible anomalies in measured DNA profiles or not? Which "population" do wo use as reference?, etc) which, as statisticians know, can be crucial to conclusions. In fact, part of a statistician's usual work when collaborating with *scientists* in any applied field, is coming to an *unfettered intuitive*, but definitely expert, *synthesis* of results of the many component statistical analyses which together form one statistical investigation. Subject matter knowledge, and knowledge of the purpose of the analysis, cannot be separated from this synthesis. This is no different from the way other experts (e.g., medical) do work, and are expected to work, in a multidisciplinary context. It is equally reproducible, equally subject to objective discussion and criticism. The problem that courts and laypersons have with understanding statistical evidence are not essentially from their problems with understanding scientific methodology and thinking in general, though of perhaps greater magnitude, due to confusion of language and the frequent (well known) failure of layman's intuition without regard to probability and uncertainty.

**Miscarriages of Justice Arising in the Context of Statistical Evidence**

Miscarriages of justice arising from issues with expert evidence are a source of grave concern for judicial officers, trial lawyers and expert witnesses alike. Their incidence is the subject of ongoing debate (see eg Naughton, 2009; Huff and Kilias, 2008; Naughton, 2007; Nobles and Schiff, 2000; Walker and Starmer, 1999). Such miscarriages have many explanations, some of them occurring because of deliberate withholding of evidence by the prosecution (Spencer, 1997), some because of poor quality scientific work, such as contamination of exhibits or unsatisfactory functioning of forensic laboratories (see eg Bromwich, 2007; Thompson, 2009; Kelly and Wearne, 1998), some because of poor cross-examination that does not expose evidentiary deficiencies (Freckelton 1997), and some because of decision-maker misevaluation of scientific data. It can be viewed as an aspect of law's struggle to integrate science and the law (see Dwyer, 2008; Faigman, 2004). Statistical evidence has played a role in a number of these cases.

Trial decisions where it is widely viewed that the wrong result was arrived at, including the Lucia de Berk saga in The Netherlands (see Derksen, 2010), in which fundamentally different forms of evidence were given in respect of the unlikelihood of a nurse being present at multiple scenes of unrelated death, provide an opportunity for professional reflection. While miscarriages of justice are tragic for those concerned and corrosive of respect for the integrity of criminal justice, they have the potential to provide an impetus for improvement in professional practice and, sometimes**,** for systemic change for the better. However, this depends upon candid identification of the sources of error and preparedness to address them. Too often this is not forthcoming in appellate decisions or even on scholarly review.

For those jurisdictions in the Anglo-American tradition that employ evidentiary admissibility thresholds, but also for those jurisdictions that depend for sound curial decision-making upon rigorous evaluation of evidence, miscarriages of justice should prompt much reflection on the part of forensic statisticians. Within the discipline there remains a tension between those who prefer the Bayesian paradigm and those who prefer the frequentist. There is no imperative within any professional group that it be homogeneous – many professions are riven by much more division of approach than forensic statistics. But it is important to acknowledge where fundamental differences lie, what gives rise to them and what consequences ensue for the application of different approaches. Transparency about such fundamental matters, even though it may be confronting, is the mark of maturity in a professional discipline (Malsch & Nijboer, 2005; Nijboer, 2011).

**Constructive Responses for Forensic Statisticians**

Identification of the parameters of disciplinary competence can encourage practitioners to be cognisant of their limitations – when they should respectfully decline to express forensic opinions and when they should stipulate qualifications upon views which may be tentative or provisional.

Issues for the Likelihood Ratio approach are the assumptions made and what was referred to in *Sienkiewicz v Greif (UK) Ltd* (2011: at [11]) as *"the adequacy of the data"*. It is incumbent upon practitioners throughout forensic science to be both rigorous in the incorporation and quantification of such assumptions and frank about any issues that might bring into question the legitimacy or the significance of such assumptions. The appropriateness of the basis of opinions led in part to the disinclination of the English Court of Appeal in *Doheny & Adams v The Queen* (1997) to accept Bayesian reasoning and it played some role also in *R v T* (2010). The key considerations are the twin pillars that induce judicial confidence in expert evidence: transparency and accountability. But other factors were at play in *R v T* and also in the Supreme Court of the United Kingdom decision in relation to epidemiology evidence in *Sienkiewicz v Greif (UK) Inc*. Courts, too have wrestled with "coincidence" evidence (see eg Brenner, 2010) and how they should factor into their reasoning processes evidence about the relative unlikelihood of unusual events occurring (see eg the de Berk case) which may in the individual case be explicable (eg SIDS cases in the one family: see *R v Clark*, 2003*; Cannings v The Queen*, 2004; *R v Matthey*, 2007)) but in the aggregate be such as to engender extreme levels of concern about the conduct of a defendant and therefore lead to prejudiced and flawed processes of reasoning. For the law this is an area of considerable sensitivity, whose underlying principles are still being worked through in difficult cases which have high personal stakes for the defendants involved. Thus legal theory itself, assisted by philosophical and analytical rigour, as well as by forensic statistics, has a role to play in enhancing decision-making and reducing the potential for miscarriages of justice.

It is apparent that forensic statistics evidence enters something of an adverse curial environment in many countries and as a discipline must proceed with both circumspection and candour. It has the potential to enhance decision-making because of its provision of a reliable language of analysis and rigorous principles of reasoning. However, instances where it proceeds on the basis of questionable hypotheses or assumptions run the risk of undermining the confidence of the courts (Nijboer, 2008).

It is trite to observe that the contributions of the discipline to notorious miscarriages of justice have detracted from its stature in the forensic domain (Broeders, 2003). What matters most for now is how the discipline responds in terms of its articulation of values, methodologies and approaches.

The future of forensic statistics as a discipline drawn upon and trusted by the courts to provide assistance that is not more prejudicial than probative (see, e.g., United States Federal Rules of Evidence, Rule 403[4]) depends on both short-term and long-term work by the profession. A substantial contribution has been made by the Royal Statistical Society (Aitken, Roberts and Jackson, 2010; see also National Academy of Sciences, 2009), in its publication about the contributions able to be made by the discipline. But more is necessary. Clear enunciation of the differences of approach within the discipline and the parameters of competence within the profession to assist legal decision-making will be important (Meintjes-Van der Walt, 2000) . This may be by way of statements of consensus or by way of codes of practice for forensic statisticians. Contrasting examples in this regard are the short-form code of the Expert Witnesses Institute (Expert Witnesses Institute, 2005), the Second Consultation Draft Codes of Practice and Conduct for Forensic Science Providers and Practitioners in the Criminal Justice System by the Forensic Science Regulator (Forensic Science Regulator, 2010), and the Code of Ethics of the Australian and New Zealand Forensic Science Society (Forensic Science Society, 2011). Such proclamations will go a significant way to address conduct such as that of Professor Sir Roy Meadow in the Sally Clark trial (see Freckelton, 2007, Freckelton and Selby, 2009) because they will place a clear onus upon those who propose to depart in their reports and oral evidence from such methodologies to justify their stance.

The creation of bodies such as a National Institute of Forensic Science, as recommended by the National Research Council of the National Academies (2009) in the United States, and as exists in Australia (National Institute of Forensic Science, 2011) and in the United Kingdom under another name, the Forensic Science Regulator, has the potential also to be constructive in terms of enhancing standards of practice in relation to forensic science generally. However, there is much to be said for discipline-specific initiatives to clarify professional issues (including those which are or may be seen to be internally divisive) and to stipulate expected standards of practice within forensic statistics.

In addition, broadening of the curriculum of law students at undergraduate level to introduce the perspectives of the sciences, including forensic statistics, is important, as is the development of postgraduate courses that explore such issues in depth. Continuing provision of ongoing professional education to practising lawyers and to judicial officers by respected members of the profession of forensic statistics has the potential to desensitise to numeracy phobias, to initiate a culture of recognition and potential acceptance of quantitative evidence, and in due course to facilitate rigorous and enhanced decision-making and evaluation by advocates and judicial decision-makers alike.

---

[4] "Although relevant, evidence may be excluded if its probative value is substantially outweighed by the danger of unfair prejudice, confusion of the issues, or misleading the jury, or by considerations of undue delay, waste of time, or needless presentation of cumulative evidence."

# References

*Adams* [1996] 2 Cr App R 467

Aitken C, Roberts P and Jackson G (2010), *Communicating and Interpreting Statistical Evidence in the Administration of Criminal Justice* (Royal Statistical Society: London): http://www.rss.org.uk/uploadedfiles/userfiles/files/Aitken-Roberts-Jackson-Practitioner-Guide-1-WEB.pdf

Amendments presented to Congress of the Federal Rules of Evidence: http://federalevidence.com/pdf/2011/04-Apr/FRE.Restyling.4-26-11SCT.pdf

Association of Forensic Science Providers, "Standards for the formulation of evaluative forensic science expert opinion". Science and Justice 49: 161–164.

Brenner C (2010). "Fundamental Problem of Forensic Mathematics – The Evidential Value of a Rare Haplotype" *Forensic Sci. Int. Genet.* 4: 281–291.

Broeders, A.P.A. (2003), Op zoek naar de bron -

Bromwich MR (2007), *Final Report of the Independent Investigator for the Houston Police Department Crime Laboratory and Property Room*: http://www.hpdlabinvestigation.org/reports/070613report.pdf

*Cannings v The Queen* [2004] EWCA Crim 1, [2004] 1 All ER 725

*Dasreef Pty Ltd v Hawchar* [2011] HCA 21

Derksen T (2010). *Lucia de Berk* (Arbeiderspers).

Doheny and Adams v The Queen [1997] 1 Cr App R 369

Dwyer D (2008). *Judicial Assessment of Expert Evidence* (Cambridge University Press, Cambridge).

De Roos Th A & Nijboer J F (2011), **"**Wrongly convicted – how the Dutch deal with their miscarriages of justice**"**, *Criminal Law Forum* 2011 (in print)

Expert Witnesses Institute (2005), *"Code of Practice for Experts"*: http://www.ewi.org.uk/files/the%20law%20and%20you/CodeofPractice.pdf

Faigman D (2004). Laboratory of Justice: The Supreme Court's 200 year Struggle to Integrate Science and the Law (Times Books, New York).

Forensic Science Regulator (2010) Second Consultation Draft Codes of Practice and Conduct for Forensic Science Providers and Practitioners in the Criminal Justice System by the Forensic Science Regulator: http://www.homeoffice.gov.uk/publications/police/forensic-science-regulator1/quality-standards-codes-practice?view=Binary

Forensic Science Society of Australia and New Zealand (2011), "Code of Ethics":

*Jones v Kaney* [2011] UKSC 13

Freckelton I (1997). "Wizards in the Crucible: Making the Boffins Accountable" in JF Nijboer and JM Reintjes (ed), *Proceedings of the First World Conference on New Trends in Criminal Investigation and Evidence* (Koninklijke Vermande, Lelystad).

Freckelton I (2006). "Immunity for Experts from Disciplinary Regulation". *Journal of Law and Medicine*. 13: 393-397.

Freckelton I and Selby H (2009). Expert Evidence: Law, Practice, Procedure and Advocacy (Thomson, Sydney).

Freckelton I (2011), "Forensic Statistics Evidence" in I Freckelton and H Selby (ed), *Expert Evidence* , looseleaf service (Sydney: Thomson Reuters).

Goldberg R (2011) *Perspectives on Causation* (Hart Publishing, Oxford).

Hodgson DH Probability and Proof in Legal Fact Reasoning" (1995) 69 ALJ 731

Huff CR and Killias M (2008). Wrongful Conviction: International Perspectives on Miscarriages of Justice (Temple University press, Philadelphia). Kelly JF and Wearne PK (1998), Tainting Evidence: Inside the Scandals at the FBI Crime Lab (The Free Press, New York).

Law Commission of England and Wales (2011): *Expert Evidence in Criminal Proceedings*: http://www.justice.gov.uk/lawcommission/docs/lc325_Expert_Evidence_Report.pdf

Malsch, M & Nijboer, J F (1999), Complex cases – Perspectives on the Netherlands criminal justice system, Amsterdam: Thela Thesis 1999

Malsch, M & Nijboer, J F (2005), *De zichtbaarheid van het recht*, Deventer: Kluwer 2005

*Makita (Australia) Pty Ltd v Sprowles* [2001] NSWCA 305

*Meadow v General Medical Council* [2006] EWCA Civ 1390, [2007] QB 462

Naughton M (2007). *Rethinking Miscarriages of Justice* (Palgrace MacMillan, London).

National Academy of Sciences (2009). *Strengthening Forensic Science in the United States: A Path Forward* (National Academies Press, Washington DC).

National Institute of Forensic Science (2011). "Vision and Mission": http://www.nifs.com.au/NIFS/NIFS_frame.html?Goals.asp&1

Naughton M (2007). Rethinking Miscarriages of Justice: Beyond The Tip of the Iceberg (Palgrave MacMillan, London).

Naughton M (2009). The Criminal Cases Review Commission: Hope for the Innocent?(Palgrave MacMillan, London)

Nijboer, J.F., De (straf)rechtspleging als lerend system, Deventer: Kluwer 2008

Nijboer J F, *Strafrechtelijk bewijsrecht*, Nijmegen: Ars Aequi Libri 2011 (6th edition)

Nobles R and Schiff D (2000), *Understanding Miscarriages of Justice* (oxford University Press, Oxford).

Redmayne, M, Roberts P, Aitken, C, Jackson G (2011) Forensic Science Evidence in Question. *Criminal Law Review*. 347-356.

*R v Clark (No 2)* [2003] EWCA Crim 1020

*R v GK* [2001] NSWCCA 41

*R v Matthey* (2007) 177 A Crim R 470; [2007] VSC 398

*R v South* [2011] EWCA Crim 754

*R v T* [2010] EWCA 2439: http://www.bailii.org/ew/cases/EWCA/Crim/2010/2439.pdf
Royal Commission on Criminal Justice (1993) ….

*Sienkiewicz v Greif (UK) Ltd* [2011] UKSC 10: http://www.bailii.org/uk/cases/UKSC/2011/10.html

Spencer J (1997). "Disclosure of Evidence in England and Wales: Progress or Regression?" " in JF Nijboer and JM Reintjes (ed), Proceedings of the First World Conference on New Trends in Criminal Investigation and Evidence (Koninklijke Vermande, Lelystad).

Thompson WC and Schumann EL (1987). Interpretation of Statistical Evidence in Criminal Trials: The Prosecutor's Fallacy and the Defense Attorney's Fallacy". Law and Human Behavior. 11(3): 167

Thompson WC (2009), "Beyond Bad Apples: Analyzing the Role of Forensic Science in Wrongful Convictions". Southwestern University Law Review. 37: 971-994.

Vosk T (November 2010), "Trial by Numbers". Champion Magazine 48: http://www.nacdl.org/public.nsf/698c98dd101a846085256eb400500c01/d3d8d819eb0b64d685257822007087ca?OpenDocument&Highlight=0,forensic,forensics,evidence

Walker K and Starmer C (1999). *Miscarriages of Justice: A Review of Justice in Error* (Blackstone Press, London).